# Posterior Probabilities for Lorenz and Stochastic Dominance of Australian Income Distributions

David Gunawan
*University of Wollongong*

William E. Griffiths
*University of Melbourne*

Duangkamon Chotikapanich
*Monash University*

8 April 2021

**Abstract**

Using HILDA data for the years 2001, 2006, 2010, 2014 and 2017, we compute posterior probabilities for dominance for all pairwise comparisons of income distributions in these years. The dominance criteria considered are Lorenz dominance and first and second order stochastic dominance. The income distributions are estimated using an infinite mixture of gamma density functions, with posterior probabilities computed as the proportion of Markov chain Monte Carlo draws that satisfy the inequalities that define the dominance criteria. We find welfare improvements from 2001 to 2006 and qualified improvements from 2006 to the later three years. Evidence of an ordering between 2010, 2014 and 2017 cannot be established.

## 1.    Introduction

Concern over the welfare implications of growth and inequality has led to numerous studies of the Australian income distribution and its components and how they have changed over time. The focus has generally been on inequality: its level, how it has changed, the relevant variable for measuring it, its drivers, and the source of data used to measure it. Alternative data sources include various surveys conducted by the Australian Bureau of Statistics (ABS), the Household, Income and Labour Dynamics in Australia (HILDA) survey[1], and tax records. Wilkins (2015) describes the nature of these data sources, and their strengths and weaknesses for producing estimates of income inequality. In Wilkins (2014) he uses a comparison with results from HILDA data to analyze the impact on inequality estimates of changes in the ABS survey methodology. Further comparisons are made in Burkauser *et al*. (2015). They examine how changes in tax law, particularly changes relating to the taxing of capital gains, can have an impact on inequality estimates derived from tax data, such as those in Atkinson and Leigh (2007). Earnings inequality, sources of its changes, and its impact on income inequality, are investigated by Borland and Coelli (2016) using data from three ABS surveys: Income Distribution Surveys, Labour Force Surveys and Employer Surveys. They consider the effect of occupational structure on earnings inequality in Coelli and Borland (2016). Sila and Dugain (2019) use data from the HILDA survey to describe changes in income, wealth and earnings inequality. Chatterjee *et al*. (2016) characterize wage inequality over an individual's life cycle using HILDA data, and decompose residual inequality into permanent and transitory components. Their model is extended by Kaplan *et al*. (2018) who, in addition to using HILDA data to decompose residual income inequality into permanent and transitory components, use ABS household expenditure survey data to examine trends in consumption and inequality, paying particular attention to the role of imputed rent. The impact of the tax-benefit system on inequality is estimated by Herault and Azpitarte (2015) using ABS data. This issue is also considered by Li *et al*. (2021) who use counterfactual distributions, estimated from HILDA data, to decompose inequality differences into those attributable to tax and welfare policies, demographic effects, market effects and residual effects. These are some recent examples of the many studies which have appeared in the literature; others, particularly earlier studies, can be found from these sources.[2]

---

[1]  The HILDA Project was initiated and is funded by the Australian Government Department of Social Services (DSS) and is managed by the Melbourne Institute of Applied Economic and Social Research (Melbourne Institute) (Watson and Wooden, 2012). The findings and views reported in this paper, however, are those of the authors and should not be attributed to either DSS or the Melbourne Institute.

Our study is a departure from the existing literature in three main ways. First, we are not just interested in inequality, but also in the more general concept of welfare. We provide evidence on whether Australia has experienced welfare improvements over the period $2001 - 2017$ by selecting four years (2001, 2006, 2010 and 2017), and using stochastic dominance to compare income distributions estimated for each of these years. Stochastic dominance is related to welfare changes for broad classes of social welfare functions. First-order stochastic dominance implies greater utility for all social welfare functions that are strictly increasing. It implies the level of income from the dominant distribution is greater than or equal to the level of income from the dominated distribution at all population proportions. Second-order stochastic dominance provides an unambiguous welfare ranking for the class of social welfare functions that are increasing and concave. It implies the sum of incomes below any population proportion is at least as great for the dominant distribution as it is for the dominated distribution. First-order stochastic dominance implies second-order stochastic dominance, but the converse is not true.[3]

When investigating inequality the existing literature tends to focus on single indices such as the Gini coefficient, and ratios of various percentiles of the distribution, or, in the case of tracking wage inequality over time, the variance of the log-wage. Instead, we investigate inequality in the context of Lorenz dominance. For two income distributions with the same mean income, Lorenz dominance implies greater utility with respect to all strictly increasing and concave social welfare functions. Generalized Lorenz dominance, where Lorenz curves are multiplied by mean income, is equivalent to second-order stochastic dominance.

Our third departure from the literature is the way in which we present evidence about dominance. An existing study, applied to Australian income distributions estimated with ABS data, is the work of Valenzuela *et al*. (2014), hereafter VLA. They use sampling theory tests proposed by Barrett and Donald (2003) to test for first and second order stochastic dominance, comparing distributions for gross and disposable income, and two expenditure categories, over the years 1983 to 2010. Their evidence for or against dominance is presented as failure to reject, or rejection, of a null hypothesis of dominance. We follow Lander *et al.* (2020), hereafter LGGC, and use a Bayesian approach to provide posterior probabilities for dominance in either direction, and the probability of no dominance.

---

[2] There is also a vast literature on changes in come distributions in other countries, and globally, and on the best way to measure and capture those income distributions. A recent example is Larrimore et al. (2021) who examine changes in U.S. income distributions using improvements to data provided by tax records.

[3] We restrict our analyses to first- and second-order stochastic dominance. However, third-order stochastic dominance, relevant for ordering distributions for social welfare functions where the marginal utility function is positive, decreasing, and strictly convex, can also be considered. See, for example, Chakravarty (2009) and Le Breton and Peluso (2009).

Assessing Lorenz or stochastic dominance involves comparing the complete range of two curves: Lorenz curves for Lorenz dominance, generalized Lorenz curves for second-order stochastic dominance, and distribution functions for first-order stochastic dominance. Because the curves, or points on those curves, are estimated from sample data, they are subject to sampling error. The existence of sampling error has led to a multitude of nonparametric statistical tests designed to test whether the difference between two curves is "significant". These tests consider estimates at a number of points; some examine the maximum distance between curves at these points, others use the joint distribution over a number of points; some use dominance as the null hypothesis, others use no dominance as the null. An extensive list of such studies, one of which is the Barrett and Donald (2003) proposal used by VLA, and another is the test for Lorenz dominance proposed by Barrett et al. (2014), is referenced in LGGC.

A major difference between our Bayesian approach and the nonparametric sampling theory approach adopted by VLA and others is that we do not specify null and alternative hypotheses. Instead, we find three posterior probabilities, the probability of dominance in each direction, and the probability of no dominance. These probabilities are estimated using the proportion of Markov chain Monte Carlo (MCMC) parameter draws, obtained from estimating each distribution as a mixture of gamma densities, for which one curve is greater than another over the range of population proportions. To ensure an accurate calculation over this range we consider a fine grid of population proportions, from 0.001 to 0.999, at 0.001 intervals. Using this approach, it is also possible to focus on some segments of the population, like the poor, if they are deemed to have particular importance. As explained in LGGC, providing three posterior probabilities is more informative than the outcomes of many sampling theory tests. A dilemma faced by sampling-theory tests is well illustrated by considering the results in VLA.

Suppose we have two distributions $X$ and $Y$. VLA perform two tests, one where the null hypothesis is that $X$ dominates $Y$ and one where the null hypothesis is that $Y$ dominates $X$. They conclude that $X$ dominates $Y$ (say) if the first null hypothesis is not rejected, and the second is rejected. However, non-rejection of a null hypothesis does not imply the null hypothesis is true; only that there is not strong evidence against the null. The proper conclusion is that either $X$ dominates $Y$ *or* neither distribution is dominant. LGGC provide examples of where the VLA strategy would lead to a conclusion that one distribution is first-order stochastically dominant even when the empirical distribution functions cross. In this case, intuition suggests dominance in neither direction is a more likely scenario; and it is likely to be picked up by a high posterior probability of no dominance. A second situation where ambiguity arises is when both null hypotheses, $X$ dominates $Y$ and $Y$ dominates $X$, cannot be rejected. It is not possible to accept both null hypotheses; that would imply $X$ dominates $Y$ *and* $Y$ dominates $X$. Faced with this awkward outcome, VLA record this result as "insignificant". It occurs when two curves are close together relative to the standard deviation of their difference. In this instance, one curve may dominate another, but because they are close together, dominance cannot be rejected. Alternatively, the curves may cross but remain close together, preventing rejection of a null hypothesis. In the first of these two cases, the posterior probability of dominance is likely to be slightly greater than 0.5; it would be close to zero if the curves cross.

We are not the first set of authors to recognize that dominance can never be established when using a null hypothesis that one distribution dominates another. To overcome this problem, Davidson and Duclos (2013) propose testing a null hypothesis of non-dominance, and devise a test based on an empirical likelihood test statistic. In this case, rejection of the null leads to a legitimate claim of dominance. However, they go on to show that, with continuous distributions, a null-hypothesis of non-dominance can never be rejected unless the range of income is restricted. The problem lies in the tails, where all quantile functions converge to zero or one. Faced with this dilemma, Davidson and Duclos argue that the only empirically sensible approach to take is to test for "restricted dominance", with the null of non-dominance and the alternative of dominance specified over a restricted range of income. By considering population proportions from 0.001 to 0.999, we are implicitly considering restricted dominance. The poorest 0.1% and the richest 0.1% of the population are being ignored. Not surprisingly, our results are sensitive to this choice. However, as we shall see, it is easy to quantify how the posterior probability of dominance changes for different ranges of population proportions.

In their work, LGGC use a finite mixture of gamma densities to estimate their income distributions. From these estimates, the corresponding Lorenz curves, generalized Lorenz curves, and distribution functions are obtained. While finite mixtures are very flexible specifications, known to approximate well a wide variety of possible density functions, they are nevertheless open to the criticism that the parametric specification might be too restrictive. To minimize this possibility we estimate an infinite mixture of gamma densities, known in the Bayesian literature as nonparametric density estimation, reflecting the fact that infinite mixtures can approximate any distribution. See, for example, Gelman *et al*. (2014, Ch.23). In addition to extending the finite gamma mixture in LGGC to an infinite gamma mixture, in the current paper we combine the MCMC algorithm for estimating the infinite gamma mixture with an algorithm that accommodates the sampling weights provided with the HILDA data. The Bayesian bootstrap is used to generate pseudo random samples using an algorithm that is described in Dong *et al.* (2014) and evaluated further in Gunawan *et al*. (2020).

The format of the paper is as follows. In Section 2 we summarize the conditions for Lorenz and first and second order stochastic dominance. Estimation of the infinite mixture of gamma densities is discussed in Section 3, with details provided in an online appendix. In Section 4 we describe how the MCMC draws are used to estimate dominance probabilities. The data are described in Section 5 and results presented in Section 6. Some concluding remarks are made in Section 7.

## 2.    Dominance Conditions

Consider an income distribution $X$ with mean $\mu_X$, density function $f_X(x)$, distribution function $F_X(x) = \int_0^x f_X(t)dt$, and first moment distribution function $F_X^{(1)} = (1/\mu_X)\int_0^x t f_X(t)dt$. For a given proportion of the population $u$, a convenient way to write the corresponding Lorenz curve is

$$L_X(u) = F_X^{(1)}\left[F_X^{-1}(u)\right] \qquad\qquad 0 \le u \le 1 \qquad\qquad (1)$$

where $F_X^{-1}(u)$ is the quantile function for $X$. With these definitions, and corresponding ones for another distribution $Y$, we can write the required dominance conditions as follows:

<u>Lorenz dominance</u> $\left( X \geq_{LD} Y \right)$

$$L_X(u) \geq L_Y(u) \quad \text{for all } 0 \leq u \leq 1 \text{ and } L_X(u) > L_Y(u) \text{ for some } 0 < u < 1 \tag{2}$$

<u>Second-order stochastic dominance</u> $\left( X \geq_{SSD} Y \right)$

$$\mu_X L_X(u) \geq \mu_Y L_Y(u) \qquad \text{for all } 0 \leq u \leq 1 \text{ and } \mu_X L_X(u) > \mu_Y L_Y(u) \text{ for some } 0 < u < 1 \tag{3}$$

<u>First-order stochastic dominance</u> $\left( X \geq_{FSD} Y \right)$

$$F_X^{-1}(u) \geq F_Y^{-1}(u) \quad \text{for all } 0 \leq u \leq 1 \text{ and } F_X^{-1}(u) > F_Y^{-1}(u) \text{ for some } 0 < u < 1 \tag{4}$$

First-order stochastic dominance can also be stated in terms of distribution functions instead of quantile functions. Another representation of second-order stochastic dominance is in terms of integrals of distribution functions. However, (2), (3) and (4) are convenient for our purpose because the range of the argument $u$ (the proportion of population) must lie within the (0,1) interval.[4]

## 3.      Estimating an Infinite Mixture of Gamma Densities

For estimating the income distributions using an infinite mixture of gamma densities, we employ the Dirichlet process mixture model proposed by Escobar and West (1995) in the context of normal distributions.[5,6] An infinite mixture of gamma densities can be written as

$$p(y \mid \mathbf{\mu}, \mathbf{v}, \mathbf{w}) = \sum_{k=1}^{\infty} w_k G\left( y \mid v_k, v_k / \mu_k \right) \tag{5}$$

where $y$ is a random income draw from the probability density function (pdf) $p(y \mid \mathbf{\mu}, \mathbf{v}, \mathbf{w})$, with parameter vectors $\mathbf{w}' = (w_1, w_2, \ldots)$, $\mathbf{\mu}' = (\mu_1, \mu_2, \ldots)$, and $\mathbf{v}' = (v_1, v_2, \ldots)$. The pdf $G\left( y \mid v_k, v_k / \mu_k \right)$ is a gamma density with mean $\mu_k > 0$ and shape parameter $v_k > 0$,

---

[4] Stating the conditions for dominance in terms of quantile functions with argument $u$ is sometimes referred to as indirect stochastic dominance, with the equivalent conditions, stated in terms of distribution functions with income as the argument, called direct stochastic dominance. Further details can be found, for example, in Davidson and Duclos (2000), Lambert (2001) and Maasoumi (1997).

[5] A similar model, using an infinite mixture of lognormal distributions to estimate income distributions, was used by Hasegawa and Kozumi (2003).

[6] Hajargasht et al. (2012) consider estimation of a wide range of parametric distributions from grouped data.

$$G\left(y \mid v_k, v_k / \mu_k\right) = \frac{\left(v_k / \mu_k\right)^{v_k}}{\Gamma\left(v_k\right)} y^{v_k - 1} \exp\left(-\frac{v_k}{\mu_k} y\right) \tag{6}$$

A description of the estimation procedure and the MCMC algorithm used to obtained the parameter draws are provided in an online appendix. For each of $m = 1, 2, \ldots, M$ MCMC parameter draws, we obtain the values

$$\boldsymbol{w}^{(m)} = \left(w_1^{(m)}, w_2^{(m)}, \ldots, w_{K^{(m)}+1}^{(m)}\right)'$$

$$\boldsymbol{v}^{(m)} = \left(v_1^{(m)}, v_2^{(m)}, \ldots, v_{K^{(m)}+1}^{(m)}\right)'$$

$$\boldsymbol{\mu}^{(m)} = \left(\mu_1^{(m)}, \mu_2^{(m)}, \ldots, \mu_{K^{(m)}+1}^{(m)}\right)'$$

The algorithm is such that, at each iteration, the infinite number of weights $w_1, w_2, \ldots$ is truncated at a point $K^{(m)}$, leaving a residual weight $w_{K^{(m)}+1}^{(m)} = 1 - \sum_{k=0}^{K^{(m)}} w_k^{(m)}$. Posterior sampling for $\boldsymbol{v}$ and $\boldsymbol{\mu}$ continues with these $K^{(m)} + 1$ weights. Then, for each parameter draw, there is a corresponding distribution function

$$F(y / \boldsymbol{\mu}^{(m)}, \boldsymbol{v}^{(m)}, \boldsymbol{w}^{(m)}) = \sum_{k=1}^{K^{(m)}+1} w_k^{(m)} F\left(y \mid v_k^{(m)}, v_k^{(m)} / \mu_k^{(m)}\right) \qquad m = 1, 2, \ldots, M \tag{7}$$

and first moment distribution

$$F^{(1)}(y / \boldsymbol{\mu}^{(m)}, \boldsymbol{v}^{(m)}, \boldsymbol{w}^{(m)}) = \frac{1}{\mu^{(m)}} \sum_{k=1}^{K_m+1} w_k^{(m)} \mu_k^{(m)} F_k^{(m)}\left(y \mid v_k^{(m)} + 1, v_k^{(m)} / \mu_k^{(m)}\right) \qquad m = 1, 2, \ldots, M \tag{8}$$

where $F_k\left(\cdot \mid v_k^{(m)} + 1, v_k^{(m)} / \mu_k^{(m)}\right)$ is the distribution function of a gamma density with parameters $\left(v_k^{(m)} + 1\right)$ and $\left(v_k^{(m)} / \mu_k^{(m)}\right)$, and $\mu^{(m)} = \sum_{k=1}^{K^{(m)}+1} w_k^{(m)} \mu_k^{(m)}$. To obtain the first-order stochastic dominance comparison in equation (4), and to find the Lorenz curve in equation (1), the distribution function in (7) needs to be inverted to obtained its corresponding quantile function. For this purpose we used an algorithm proposed in the technical appendix to LGGC.

## 4. Estimating Posterior Probabilities of Dominance

Having obtained MCMC draws on complete parameter vectors from two distributions, $\boldsymbol{\theta}_X = (\boldsymbol{w}_X, \boldsymbol{v}_X, \boldsymbol{\mu}_X)$ and $\boldsymbol{\theta}_Y = (\boldsymbol{w}_Y, \boldsymbol{v}_Y, \boldsymbol{\mu}_Y)$, we are in a position to compute corresponding values for Lorenz curves $L_X\left(u; \boldsymbol{\theta}_X\right)$ and $L_Y\left(u; \boldsymbol{\theta}_Y\right)$, generalized Lorenz curves $\mu_X L_X\left(u; \boldsymbol{\theta}_X\right)$ and $\mu_Y L_Y\left(u; \boldsymbol{\theta}_Y\right)$, and quantile functions $F_X^{-1}(u;\boldsymbol{\theta}_X)$ and $F_Y^{-1}(u;\boldsymbol{\theta}_Y)$, for a grid of $u$ values. Let $C_X(u;\boldsymbol{\theta}_X)$ and $C_Y(u;\boldsymbol{\theta}_Y)$ be generic curves, representing any pair of the three curves whose dominance properties are being considered. Also, let $C_X(u;\boldsymbol{\theta}_X^{(m)})$ and $C_Y(u;\boldsymbol{\theta}_Y^{(m)})$, $m=1,2,\ldots,M$ be their values at each of the $M$ MCMC draws, for a given population proportion $u$. We consider values for $u$ from 0.001 to 0.999 at intervals of 0.001. Following LGGC, we estimate the posterior probability of $X$ dominating $Y$ as the proportion of draws for which $C_X(u;\boldsymbol{\theta}_X^{(m)}) \geq C_Y(u;\boldsymbol{\theta}_Y^{(m)})$ for all $u$. To express this proportion mathematically, let $u_i = 0.001i$ and let $I[\cdot]$ denote an indicator function equal to 1 if its argument is true and zero, otherwise. Then we have,

$$P(X \text{ dominates } Y) = \frac{1}{M} \sum_{m=1}^{M} \prod_{i=1}^{999} I\left[ C_X(u_i;\boldsymbol{\theta}_X^{(m)}) \geq C_Y(u_i;\boldsymbol{\theta}_Y^{(m)}) \right]$$

$$P(Y \text{ dominates } X) = \frac{1}{M} \sum_{m=1}^{M} \prod_{i=1}^{999} I\left[ C_Y(u_i;\boldsymbol{\theta}_Y^{(m)}) \geq C_X(u_i;\boldsymbol{\theta}_X^{(m)}) \right]$$

$$P(\text{neither distribution dominates}) = 1 - P(X \text{ dominates } Y) - P(Y \text{ dominates } X)$$

Since distributions $X$ and $Y$ are estimated independently, the order of $\boldsymbol{\theta}_X$ and $\boldsymbol{\theta}_Y$ in $C_X(u_i;\boldsymbol{\theta}_X^{(m)}) \geq C_Y(u_i;\boldsymbol{\theta}_Y^{(m)})$ is arbitrary. To check the results, the order of one of the vectors was randomized 1,000 times and for each randomization a dominance probability was calculated. There were no substantive changes in the probabilities. To give an indication of the variation in probabilities across different orderings, in Table 1 we report the smallest, largest and average values over the orderings for some selected pairwise comparisons: two cases each for small, intermediate and large probabilities. The average of estimates over the randomizations was taken as the final estimate of the posterior probabilities.

A by-product of the estimation procedure is a plot of the curve

$$P_{X \geq Y}(u) = \frac{1}{M} \sum_{M=1}^{M} I\left[ C_X(u_i;\boldsymbol{\theta}_X^{(m)}) \geq C_Y(u_i;\boldsymbol{\theta}_Y^{(m)}) \right]$$

against the value of $u$. Called probability curves by LGGC, these curves give the probability of "dominance" at a *given* population proportion $u$. The probability of dominance over any range of $u$ will be no greater than the minimum value of $P_{X \geq Y}(u)$ within that range. This characteristic makes $P_{X \geq Y}(u)$ a valuable device for finding the population proportions which have the greatest impact on the probability of dominance. If a dominance probability is largely determined by behaviour in the tails of the distribution, we can examine the sensitivity of the probability to omission of extreme values of $u$. Also, if we are concerned with a particular segment of the population, say the poor, we can see how the probability of dominance changes if only the poor are considered.

## 5.    Data

To access dominance over the period 2001 to 2017, data were extracted from waves of the HILDA data corresponding to years 2001, 2006, 2010, 2014 and 2017. The chosen years are at reasonably spaced intervals and are sufficient to investigate a large number of pairwise comparisons. An advantage of choosing HILDA data, relative to data from the household expenditure surveys from the ABS, is that it does not suffer from ABS changes to the definitions of its variables, as described in Wilkins (2015). A possible disadvantage is that it is less representative than the ABS surveys, especially in light of sample attrition over time. The distributions for each of the years are estimated independently; in light of the panel nature of the data, they can be viewed as marginal distributions from a multivariate joint distribution. [7] The variable chosen for welfare comparisons was household disposable income, converted to a per individual basis. Many other alternatives such as permanent income, wealth, earnings and expenditure have been considered in the literature referenced in the Introduction. To obtain net disposable income we subtracted "total disposable income negative per household" from "total disposable income positive per household". The conversion of this variable to a per individual basis involves choice of an equivalence scale and the allocation of equivalised income to members of each household. For these steps we followed Sila and Dugain (2019). To obtain equivalised income, net disposable income was divided by the square root of the number of individuals in the household. This quantity was assigned to each member of the household. Values were deflated using the Consumer Price Index, treating 2000/2001 as the base. The observational units were all individuals aged 15 and above;

---

[7] Methods for combining marginal distributions to form a joint distribution via copulas have been investigated in the context of HILDA data by Vinh *et al.* (2010).

those aged less than 15 were omitted because of the unavailability of sampling weights for these individuals. Thus, the total number of children was used in the calculation of equivalised income, but children less than 15 were not included in the samples. Incomes that were non-positive were omitted from the samples: 0.55% in 2001, 0.39% in 2006, 0.35% in 2010, 0.24% in 2014 and 0.30% in 2017. These values cannot be readily handled in our analysis, and, although the numbers are relatively small, they could have a bearing on our conclusions about dominance. We need to keep in mind that, by restricting our results to the middle 99.8% of the population with positive incomes, we are also ignoring households with non-positive incomes. For examining distributions for metropolitan versus non-metropolitan areas, major urban or major city was classified as metropolitan. All other areas were classified as non-metropolitan.

Summary statistics for the income series are presented in Table 2. Sampling weights provided with the HILDA data were used for calculating the means, standard deviations and Gini coefficients. Mean income is increasing throughout the period. Inequality as measured by the Gini coefficient is highest in 2006. However, the standard deviation is highest in 2017.

## 6.    Results

Estimates of the income densities for each year are plotted in Figure 1. The units are hundreds of 2001 dollars, equivalized according to household size. The densities are all bimodal with a sharp peak between $10,000 - $20,000 and a lesser peak in the range $20,000 - $40,000. Prior to 2010 there have been clear shifts to the right, but the distributions for 2010, 2014 and 2017 are similar, particularly those in 2014 and 2017. The posterior means and standard deviations for mean income and the Gini coefficient are reported in Table 3. It is reassuring that these values are close to the values from the raw data, reported in Table 2. The mean income estimates are in line with our observations about the density functions: relatively large increases from 2001 to 2006 and from 2006 to 2010, and relatively small increases thereafter. The posterior densities for mean incomes and the Gini coefficients are plotted in Figures 2 and 3, respectively. The relative closeness of the mean incomes for the last three periods compared to those for the earlier two periods is also reflected in the posteriors in Figure 2. The 2001 posterior for the Gini coefficient suggests a lower level of inequality in that year compared to the others. There is considerable overlap in the densities in the other years.

*6.1 First-order stochastic dominance*

A more complete picture of whether or not there has been a welfare improvement is obtained by comparing distribution functions and (generalized) Lorenz curves, and computing their dominance probabilities. The distribution functions plotted in Figure 4 suggest that distributions in 2010, 2014 and 2017 all FSD the 2001 and 2006 distributions, with no clear ranking between 2010, 2014 and 2017. The dominance probabilities in Table 4 confirm the likely dominance of 2001 by all other years – the probabilities are all greater than 0.91. However, in all other pairwise comparisons, the probabilities for no dominance are all greater than 0.79. From examining Figure 4, this result is expected for comparisons between 2010, 2014 and 2017, but it is surprising that there was not more evidence of dominance of 2006 by these three years. Examining the probability curves we discover that the low probabilities for dominance of 2006 by the later years can be attributed to behaviour in the tails of the distributions. As an example, in Figure 5, we plot the probability curve for 2017 FSD 2006. The low probability of dominance of 0.0051 comes from the left tails of the distributions. The upper tails also reduce the probability, although not to the same extent. If the tails are ignored by changing the range over which dominance is considered from $0.001 \le u \le 0.999$ to $0.04 \le u \le 0.96$, the dominance probability becomes one. This outcome reinforces the argument made by Davidson and Duclos (2013), that, because all points coverage at the end points, it is inevitable that dominance can only be established over a restricted range.

*6.2 Generalized Lorenz dominance*

For GLD, we compare the generalized Lorenz curves plotted in Figure 6. Visually, we expect to find a relationship similar to that expected from Figure 4, namely, that 2001 is dominated by 2006 which in turn is dominated by 2010, 2014 and 2017, with very little difference between the last three years. Noting that the dominance probabilities for GLD will always be at least as great as those for FSD, in Table 5 we observe that the probabilities for dominance over 2001 are always greater than 0.96. It is again true that 2006 is not dominated by any of the later years; the probabilities of no dominance are all greater than 0.97. Checking the probability curves, we again find the issue is in the tails, but not both tails as was the case for FSD. In this case the left tails are the source of no dominance; an example is given in Figure 7 where the probability curve for 2017 GLD 2006 is plotted. Most of the population are better off in the later year, but the poor people are not. Restricting the range to $0.04 \leq u \leq 0.96$ yields a dominance probability of 0.8918, a value less than the FSD probability of one. When a bottom segment of the population is ignored, it is no longer true that a GLD probability will necessarily be at least as great as the corresponding FSD probability unless the distribution is truncated at the lower bound and normalized accordingly.

Another point worth noting from Table 5 that is not evident from Figure 6 is that the probabilities of dominance for pairwise comparisons of the later three years are no longer negligible. No dominance is still the most likely outcome, but with $\Pr\left(2014 \geq_{GLD} 2010\right) = 0.25$ and $\Pr\left(2017 \geq_{GLD} 2014\right) = 0.21$, dominance is still a possibility. A plot of the probability curve for 2017 GLD 2014 is given in Figure 8. It reveals a relatively high probability over most population proportions. What is perhaps surprising is the overall dominance probability of 0.21, when the minimum value of the probability curve is greater than 0.4. It can be explained by the nature of the curve. It is not monotonic. It increases, decreases, increases, decreases and then increases again. If each local minimum introduces a new set of MCMC draws that do not satisfy the required inequality, then the probability of dominance can fall well below the global minimum of the curve. Restricting the range to $0.3 \leq u \leq 0.97$ led to a dominance probability of 0.73, a value close to the global minimum within this range.

*6.3 Lorenz dominance*

The Lorenz curves plotted in Figure 9, and the posterior densities for the Gini coefficient in Figure 3, suggest that inequality is less in 2001 than it is in the other years, and that it is difficult to separate the other three years. However, from Table 6 we find that no dominance is the most likely outcome for all pairwise comparisons. The probabilities of no dominance are all greater than 0.87. To illustrate the behaviour, in Figure 10 we plot the probability curves for 2001 LD 2017 and 2017 LD 2001. Because these curves are mirror images, we can say that 2001 fails to dominate 2017 because of behaviour in the left tails of the Lorenz curves, or, 2017 fails to Lorenz dominate 2001 because of behaviour in the right tails of the Lorenz curves.

*6.4 Dominance for the poorest 10%*

In Tables 7, 8 and 9 we report dominance probabilities for the poorest 10% of the population. For FSD (Table 7) and GLD (Table 8) the results are generally in line with those for the whole population in the sense that 2001 is dominated by all subsequent years and, for all other pairwise comparisons, no dominance is the most likely outcome. There are slight increases in the probability of dominance, as one would expect, since the number of draws satisfying dominance in a restricted range must be at least as great as the number over the complete range.

Lorenz dominance in this context implies that all population proportions up to 0.1 are getting proportionately more income.[8] If an earlier distribution is dominated by a later distribution, it implies growth has been pro-poor. Whereas there was no evidence of dominance when the complete Lorenz curve was considered, in this sense there are some relatively large dominance probabilities. There is evidence that growth has been pro-poor from 2001 to 2006 and from 2001 to 2014, $\Pr\left(2006 \geq_{LD} 2001\right) = 0.76$ and $\Pr\left(2014 \geq_{LD} 2001\right) = 0.67$, but there is also some evidence that growth has not been pro-poor beyond 2006, $\Pr\left(2006 \geq_{LD} 2014\right) = 0.46$ and $\Pr\left(2006 \geq_{LD} 2017\right) = 0.71$.

*6.5 Metropolitan versus non-metropolitan subgroups*

Dominance criteria can also be used to track welfare over time for subgroups of the population and to compare subgroups at a particular point in time. To illustrate, we have chosen metropolitan and non-metropolitan subgroups. The country versus city divide often receives attention in the media, particularly from disgruntled rural politicians. Mixtures of gamma densities were estimated for both subgroups for the years 2001, 2006, 2010, 2014 and 2017. Table 10 contains the sample means, standard deviations and Gini coefficients for each subgroup and year, as well as the mean incomes and Gini coefficients estimated from the mixture of gamma densities. Examining the mean incomes suggests metropolitan incomes are substantially above non-metropolitan incomes, and that changes in incomes over time are consistent with those for the complete sample. Changes from 2001 to 2006 and from 2006 to 2010 are large relative to changes from 2010 to 2014 and from 2014 to 2017. The Gini coefficients suggest inequality increased for both groups from 2001 to 2006, but has been relatively constant thereafter. There is little difference between the metropolitan and non-metropolitan Gini coefficients in most years; 2017 is an exception with metropolitan inequality being greater in this year. It is perhaps surprising that there was not more divergence between the metropolitan and non-metropolitan Gini coefficients prior to 2017. Increasing inequality has generally been associated with a growing income share for the top incomes, a growth that is more likely in the metropolitan subgroup.

---

[8] It is not a comparison of inequality within the poorest 10% of the population.

When dominance criteria are applied to assess welfare changes over time for each subgroup, the conclusions reached for the metropolitan subgroup are similar to those reached for the overall population. However, for the non-metropolitan subgroup there are a few differences which are highlighted in Table 11. For FSD and GLD of 2014 over 2001 and 2010 over 2001 the non-metropolitan dominance probabilities are much less than those for the metropolitan subgroup.

Dominance probabilities for comparing the metropolitan and non-metropolitan subgroups in each year are displayed in Table 12. The probabilities for non-metropolitan dominance (FSD or GLD) are zero in every year. The probabilities for metropolitan dominance range from 0.1545 in 2017 to 0.6062 in 2010 for FSD, and from 0.2882 in 2017 to 0.9923 in 2014 for GLD. It is interesting that these probabilities are the highest in 2010 and 2014, the years in which it was more difficult to establish that the non-metropolitan distributions dominated the 2001 non-metropolitan distribution. We also note that, if the tails of the distributions are ignored, then the probabilities for metropolitan being dominant, for both FSD and GLD, are close to 1 in every year. We illustrate this fact by plotting the 2017 FSD and GLD probability curves in Figure 12. When the ranges for these two curves are restricted to $0.04 \leq u \leq 0.96$, the probabilities increase from 0.1545 to 0.9968 for FSD and from 0.2882 to 0.9845 for GLD.

The Lorenz dominance probabilities in Table 12 show little evidence of dominance by either subgroup. The probabilities for no dominance are all greater than 0.9. The probability curves for non-metropolitan Lorenz dominating metropolitan for all years are plotted in Figure 13. Their shapes are quite different. Most have a relatively high maximum and a relatively low minimum, suggesting the Lorenz curves cross. That for 2017 suggests the probability of metropolitan being dominant would be relatively high if the tails were ignored, particularly the right tails.

This last fact is confirmed by the LD probabilities in Table 13 where dominance probabilities for the poorest 10% are recorded. Here there are three years where the probability of LD by the non-metropolitan subgroup are more substantial: 0.3578 in 2001, 0.5739 in 2006 and 0.5323 in 2017. An examination of the tails of the probability curves in Figure 13 reveals why this increase in probabilities has occurred and why there have not been similar increases in 2010 and 2014. Table 13 also contains dominance probabilities for FSD and GLD for the poorest 10% of the metropolitan and non-metropolitan subgroups. Comparing Tables 12 and 13, we note that the absence of the right tails has led to increases in every year for the probabilities of metropolitan being FSD over non-metropolitan. However, no dominance is still the most likely outcome in 2001, 2006 and 2017, and FSD continues to be the most likely outcome in 2010 and 2014. For GLD the probabilities for metropolitan dominance remain the same when only the poorest 10% are considered. The identical values can be explained by the probability curves where the minimum value is within the region $0 \le u \le 0.1$, as illustrated for 2017 in Figure 12. There are small but non-zero values for non-metropolitan dominance in 2001, 2006 and 2017. The increases from zero, obtained when the whole range of $u$ was considered in Table 12, occur because their probability curves, like that in Figure 12, have not quite reached their maximum value at $u = 0.1$.

## 7. Concluding Remarks

We have used an innovative approach to assess first and second order stochastic dominance orderings and Lorenz ordering of the Australian household income distribution at different points in time, as well as a comparison of metropolitan and non-metropolitan household income distributions. Our innovations include expressing empirical information about dominance in terms of posterior probabilities, and using Bayesian estimation of an infinite mixture of gamma densities to provide flexible fits to the income distributions.

Using either FSD or GLD as a metric, we find strong evidence that welfare in 2001 is less than in the subsequent years: 2006, 2010, 2014 and 2017. However, pairwise comparisons of the later four years show no evidence of welfare differences unless the range of population proportions is restricted away from the tails of the distributions. With such a restriction, we can conclude that 2006 is dominated by the later three years, but an ordering of 2010, 2014 and 2017 is still not possible. Making the same comparisons for the poorest 10% of the population leads to similar conclusions. The poor are worse off in 2001 than in the later years, but it is hard to discriminate between the later years.

If we are interested only in inequality and consider Lorenz dominance instead of FSD or GLD, examining the complete Lorenz curves reveals no evidence of any ordering over the four years, despite strong posterior evidence that the Gini coefficient is smaller in 2001. When a comparison is restricted to the poorest 10%, we find some evidence that their income shares were better in 2006 and 2014 than they were in 2001, but 2006 was a better year than both 2014 and 2017.

Comparing the metropolitan and non-metropolitan subgroups in each year suggests the metropolitan subgroup is better off in terms of both FSD and GLD. However, to make this conclusion with high posterior probabilities for all years, the tails had to be excluded, with population proportions restricted to $0.04 \leq u \leq 0.96$. For Lorenz dominance there is some evidence, but not strong evidence, that the non-metropolitan group was better off in 2001, 2006 and 2017.

Our study has limitations. Results from using HILDA data will not necessarily carry over to those which might be obtained using other datasets. ABS surveys are larger and likely to be more representative, despite potential issues with changes in variable definitions. It can be argued that it is more relevant to consider expenditure than income. Refinements that consider permanent income and include imputed rent (Kaplan et al., 2018), or the use of taxation data (Burkhauser et al., 2015) are other alternatives. There are also alternatives to our choice of equivalent scale, and method for allocating household income to members of the household.

Although their numbers are relatively small, it is possible that omission of the negative and zero observations may have an impact on the results. One way forward would be to place a point mass on such observations which would scale back the remainder of the distribution. Whether different scaling in different years would be sufficient to change the results is an open question.

It is reasonable to ask about the extent to which dominance results are driven by small intersections of the curves in the tails of the distributions and whether such intersections are relevant from a practical standpoint. These intersections can lead to high posterior probabilities of no dominance when other measures such as the Gini coefficient suggest more definite changes. By their nature, intersections in the tails will be small. If they are thought to be practically irrelevant, the probability curves should be examined, and the impact of considering fewer observations in the tails can be assessed.

The potential for applying the tools used in this paper is enormous. We have only scratched the surface. Possible extensions include comparing before and after tax income distributions, or the resulting distributions from other government interventions, and the development of multivariate distributions for comparing multivariate welfare functions.

TABLE 1

*Examples of Variation in Probabilities from Reordering Draws*

| Probability | Minimum Bound | Average | Maximum Bound |
|---|---|---|---|
| $\Pr\left(2014 \succeq_{GLD} 2017\right)$ | 0.0011 | 0.0028 | 0.0044 |
| $\Pr\left(2017 \succeq_{GLD} 2014\right)$ | 0.1984 | 0.2063 | 0.2149 |
| $\Pr\left(2017 \succeq_{GLD} 2010\right)$ | 0.2927 | 0.3006 | 0.3101 |
| $\Pr\left(2014 \succeq_{FSD} 2010\right)$ | 0.0689 | 0.0762 | 0.0831 |
| $\Pr\left(2017 \succeq_{FSD} 2001\right)$ | 0.9098 | 0.9160 | 0.9219 |
| $\Pr\left(2006 \succeq_{FSD} 2010\right)$ | 0.9924 | 0.9937 | 0.0051 |

*Notes*: The minimum and maximum bounds are the smallest and largest probability estimates, respectively, from 1,000 random reorderings of the MCMC draws. The averages of the 1,000 estimates are reported in the "Average" column and in later tables.

TABLE 2

*Summary Statistics for Incomes*

| | 2001 | 2006 | 2010 | 2014 | 2017 |
|---|---|---|---|---|---|
| Sample mean | 31,791 | 37,133 | 41,990 | 43,502 | 44,446 |
| Minimum | 62 | 104 | 32 | 62 | 22 |
| Maximum | 299,494 | 557,074 | 573,183 | 586,767 | 653,686 |
| Standard deviation | 20,149 | 32,960 | 31,615 | 34,845 | 37,749 |
| Gini | 0.3126 | 0.3458 | 0.3280 | 0.3349 | 0.3398 |
| Sample size | 13,838 | 12,766 | 13,374 | 17,250 | 17,271 |

*Notes*: Equivalised income is calculated as [disposable income positive per household (*ahifditp*) − disposable income negative per household (*ahifditm*)] ÷ $\sqrt{\text{persons in household (ahhpers)}}$, and then deflated with the Consumer Price Index, with 2000/2001 base. The calculations for the mean, standard deviation and Gini coefficient have been weighted using the weight series *ahhwtrps*.

TABLE 3

*Posterior Means (Standard Deviations) for Mean Income and the Gini Coefficient*

| | 2001 | 2006 | 2010 | 2014 | 2017 |
|---|---|---|---|---|---|
| Means | 316.95 (2.39) | 382.60 (4.10) | 418.67 (3.87) | 435.06 (3.79) | 443.67 (4.19) |
| Gini | 0.3132 (0.0035) | 0.3362 (0.0055) | 0.3287 (0.0047) | 0.3363 (0.0044) | 0.3410 (0.0048) |

*Notes:* The values are those from the gamma mixture model. The means are in units of hundreds of equivalized 2001 dollars.

TABLE 4

*First Order Stochastic Dominance Probabilities*

| | A 2017 | B 2014 | A 2014 | B 2010 | A 2010 | B 2006 | A 2006 | B 2001 |
|---|---|---|---|---|---|---|---|---|
| $\Pr\left(A \geq_{FSD} B\right)$ | 0.0145 | | 0.0762 | | 0.0005 | | 0.9937 | |
| $\Pr\left(B \geq_{FSD} A\right)$ | 0.0000 | | 0.0000 | | 0.0000 | | 0.0000 | |
| $\Pr\left(\text{no dominance}\right)$ | 0.9855 | | 0.9238 | | 0.9995 | | 0.0063 | |
| | | | 2017 | 2010 | 2014 | 2006 | 2010 | 2001 |
| $\Pr\left(A \geq_{FSD} B\right)$ | | | 0.2028 | | 0.0009 | | 0.9357 | |
| $\Pr\left(B \geq_{FSD} A\right)$ | | | 0.0000 | | 0.0000 | | 0.0000 | |
| $\Pr\left(\text{no dominance}\right)$ | | | 0.7972 | | 0.9991 | | 0.0643 | |
| | | | | | 2017 | 2006 | 2014 | 2001 |
| $\Pr\left(A \geq_{FSD} B\right)$ | | | | | 0.0051 | | 0.9722 | |
| $\Pr\left(B \geq_{FSD} A\right)$ | | | | | 0.0000 | | 0.0000 | |
| $\Pr\left(\text{no dominance}\right)$ | | | | | 0.9949 | | 0.0278 | |
| | | | | | | | 2017 | 2001 |
| $\Pr\left(A \geq_{FSD} B\right)$ | | | | | | | 0.9160 | |
| $\Pr\left(B \geq_{FSD} A\right)$ | | | | | | | 0.0000 | |
| $\Pr\left(\text{no dominance}\right)$ | | | | | | | 0.0840 | |

TABLE 5
*Generalized Lorenz Dominance Probabilities*

| | A 2017 | B 2014 | A 2014 | B 2010 | A 2010 | B 2006 | A 2006 | B 2001 |
|---|---|---|---|---|---|---|---|---|
| $\Pr\left(A \succeq_{GLD} B\right)$ | | 0.2063 | | 0.2478 | | 0.0285 | | 0.9959 |
| $\Pr\left(B \succeq_{GLD} A\right)$ | | 0.0028 | | 0.0000 | | 0.0000 | | 0.0000 |
| $\Pr\left(\text{no dominance}\right)$ | | 0.7909 | | 0.7522 | | 0.9715 | | 0.0041 |
| | | | 2017 | 2010 | 2014 | 2006 | 2010 | 2001 |
| $\Pr\left(A \succeq_{GLD} B\right)$ | | | | 0.3006 | | 0.0117 | | 0.9798 |
| $\Pr\left(B \succeq_{GLD} A\right)$ | | | | 0.0000 | | 0.0000 | | 0.0000 |
| $\Pr\left(\text{no dominance}\right)$ | | | | 0.6994 | | 0.9883 | | 0.0202 |
| | | | | | 2017 | 2006 | 2014 | 2001 |
| $\Pr\left(A \succeq_{GLD} B\right)$ | | | | | | 0.0223 | | 0.9884 |
| $\Pr\left(B \succeq_{GLD} A\right)$ | | | | | | 0.0000 | | 0.0000 |
| $\Pr\left(\text{no dominance}\right)$ | | | | | | 0.9777 | | 0.0116 |
| | | | | | | | 2017 | 2001 |
| $\Pr\left(A \succeq_{GLD} B\right)$ | | | | | | | | 0.9606 |
| $\Pr\left(B \succeq_{GLD} A\right)$ | | | | | | | | 0.0000 |
| $\Pr\left(\text{no dominance}\right)$ | | | | | | | | 0.0394 |

TABLE 6
*Lorenz Dominance Probabilities*

| | A 2017 | B 2014 | A 2014 | B 2010 | A 2010 | B 2006 | A 2006 | B 2001 |
|---|---|---|---|---|---|---|---|---|
| $\Pr\left(A \succeq_{LD} B\right)$ | | 0.0069 | | 0.0046 | | 0.0011 | | 0.0000 |
| $\Pr\left(B \succeq_{LD} A\right)$ | | 0.1132 | | 0.0269 | | 0.0045 | | 0.0000 |
| $\Pr\left(\text{no dominance}\right)$ | | 0.8799 | | 0.9685 | | 0.9944 | | 1.0000 |
| | | | 2017 | 2010 | 2014 | 2006 | 2010 | 2001 |
| $\Pr\left(A \succeq_{LD} B\right)$ | | | | 0.0003 | | 0.0005 | | 0.0000 |
| $\Pr\left(B \succeq_{LD} A\right)$ | | | | 0.0918 | | 0.0131 | | 0.0112 |
| $\Pr\left(\text{no dominance}\right)$ | | | | 0.9079 | | 0.9864 | | 0.9888 |
| | | | | | 2017 | 2006 | 2014 | 2001 |
| $\Pr\left(A \succeq_{LD} B\right)$ | | | | | | 0.0001 | | 0.0000 |
| $\Pr\left(B \succeq_{LD} A\right)$ | | | | | | 0.1170 | | 0.0185 |
| $\Pr\left(\text{no dominance}\right)$ | | | | | | 0.8829 | | 0.9815 |
| | | | | | | | 2017 | 2001 |
| $\Pr\left(A \succeq_{LD} B\right)$ | | | | | | | | 0.0000 |
| $\Pr\left(B \succeq_{LD} A\right)$ | | | | | | | | 0.0529 |
| $\Pr\left(\text{no dominance}\right)$ | | | | | | | | 0.9471 |

Table 7

*First Order Stochastic Dominance Probabilities for the Poorest 10% of the Population*

|  | A 2017 | B 2014 | A 2014 | B 2010 | A 2010 | B 2006 | A 2006 | B 2001 |
|---|---|---|---|---|---|---|---|---|
| $\Pr\left(A \geq_{FSD} B\right)$ | 0.1111 | | 0.1643 | | 0.0154 | | 0.9937 | |
| $\Pr\left(B \geq_{FSD} A\right)$ | 0.0543 | | 0.0000 | | 0.0005 | | 0.0000 | |
| $\Pr\left(\text{no dominance}\right)$ | 0.8946 | | 0.8357 | | 0.9841 | | 0.0.0063 | |
|  |  |  | 2017 | 2010 | 2014 | 2006 | 2010 | 2001 |
| $\Pr\left(A \geq_{FSD} B\right)$ | | | 0.2257 | | 0.0073 | | 0.939 | |
| $\Pr\left(B \geq_{FSD} A\right)$ | | | 0.0000 | | 0.0000 | | 0.0000 | |
| $\Pr\left(\text{no dominance}\right)$ | | | 0.7743 | | 0.9927 | | 0.0604 | |
|  |  |  |  |  | 2017 | 2006 | 2014 | 2001 |
| $\Pr\left(A \geq_{FSD} B\right)$ | | | | | 0.0125 | | 0.9722 | |
| $\Pr\left(B \geq_{FSD} A\right)$ | | | | | 0.0000 | | 0.0000 | |
| $\Pr\left(\text{no dominance}\right)$ | | | | | 0.9875 | | 0.0278 | |
|  |  |  |  |  |  |  | 2017 | 2001 |
| $\Pr\left(A \geq_{FSD} B\right)$ | | | | | | | 0.9162 | |
| $\Pr\left(B \geq_{FSD} A\right)$ | | | | | | | 0.0000 | |
| $\Pr\left(\text{no dominance}\right)$ | | | | | | | 0.0838 | |

Table 8

*Generalized Lorenz Dominance Probabilities for the Poorest 10% of the Population*

| | A 2017 | B 2014 | A 2014 | B 2010 | A 2010 | B 2006 | A 2006 | B 2001 |
|---|---|---|---|---|---|---|---|---|
| $\Pr(A \geq_{GLD} B)$ | | 0.2498 | | 0.2521 | | 0.0285 | | 0.9959 |
| $\Pr(B \geq_{GLD} A)$ | | 0.2194 | | 0.0046 | | 0.0189 | | 0.0000 |
| $\Pr(\text{no dominance})$ | | 0.5308 | | 0.7433 | | 0.9526 | | 0.0.0041 |

| | A 2017 | B 2010 | A 2014 | B 2006 | A 2010 | B 2001 |
|---|---|---|---|---|---|---|
| $\Pr(A \geq_{GLD} B)$ | | 0.3009 | | 0.0117 | | 0.9798 |
| $\Pr(B \geq_{GLD} A)$ | | 0.0019 | | 0.0004 | | 0.0000 |
| $\Pr(\text{no dominance})$ | | 0.6972 | | 0.9879 | | 0.0202 |

| | A 2017 | B 2006 | A 2014 | B 2001 |
|---|---|---|---|---|
| $\Pr(A \geq_{GLD} B)$ | | 0.0223 | | 0.9884 |
| $\Pr(B \geq_{GLD} A)$ | | 0.0000 | | 0.0000 |
| $\Pr(\text{no dominance})$ | | 0.9777 | | 0.0116 |

| | A 2017 | B 2001 |
|---|---|---|
| $\Pr(A \geq_{GLD} B)$ | | 0.9606 |
| $\Pr(B \geq_{GLD} A)$ | | 0.0000 |
| $\Pr(\text{no dominance})$ | | 0.0394 |

TABLE 9

*Lorenz Dominance Probabilities for the Poorest 10% of the Population*

|  | A 2017 | B 2014 | A 2014 | B 2010 | A 2010 | B 2006 | A 2006 | B 2001 |
|---|---|---|---|---|---|---|---|---|
| $\Pr\left(A \succeq_{LD} B\right)$ | 0.1327 | | 0.1814 | | 0.0020 | | 0.7631 | |
| $\Pr\left(B \succeq_{LD} A\right)$ | 0.3034 | | 0.0527 | | 0.8783 | | 0.0000 | |
| $\Pr\left(\text{no dominance}\right)$ | 0.5639 | | 0.7659 | | 0.1197 | | 0.2369 | |
|  |  |  | 2017 | 2010 | 2014 | 2006 | 2010 | 2001 |
| $\Pr\left(A \succeq_{LD} B\right)$ | | | 0.1743 | | 0.0012 | | 0.2381 | |
| $\Pr\left(B \succeq_{LD} A\right)$ | | | 0.1517 | | 0.4594 | | 0.0117 | |
| $\Pr\left(\text{no dominance}\right)$ | | | 0.6740 | | 0.5394 | | 0.7502 | |
|  |  |  |  |  | 2017 | 2006 | 2014 | 2001 |
| $\Pr\left(A \succeq_{LD} B\right)$ | | | | | 0.0011 | | 0.6662 | |
| $\Pr\left(B \succeq_{LD} A\right)$ | | | | | 0.7090 | | 0.0075 | |
| $\Pr\left(\text{no dominance}\right)$ | | | | | 0.2899 | | 0.3263 | |
|  |  |  |  |  |  |  | 2017 | 2001 |
| $\Pr\left(A \succeq_{LD} B\right)$ | | | | | | | 0.3929 | |
| $\Pr\left(B \succeq_{LD} A\right)$ | | | | | | | 0.0599 | |
| $\Pr\left(\text{no dominance}\right)$ | | | | | | | 0.5472 | |

TABLE 10

*Summary Statistics for Metropolitan and Non-metropolitan Incomes*

|  | 2001 | | 2006 | | 2010 | | 2014 | | 2017 | |
|---|---|---|---|---|---|---|---|---|---|---|
|  | Met | Non-Met | Met | Non-Met | Met | Non-Met | Met | Non-Met | Met | Non-Met |
| Sample mean | 33,491 | 27,729 | 40,599 | 32,649 | 44,338 | 35,709 | 45,846 | 37,690 | 46,848 | 38,514 |
| Estimated mean | 33,486 | 27,696 | 40,468 | 32,692 | 44,227 | 36,155 | 45,777 | 37,497 | 46,961 | 38,374 |
| Standard deviation | 20,663 | 18,232 | 33,293 | 28,034 | 33,056 | 26,892 | 35,792 | 31,637 | 39,363 | 32,347 |
| Sample Gini | 0.3074 | 0.3131 | 0.3321 | 0.3296 | 0.3236 | 0.3256 | 0.3312 | 0.3319 | 0.3403 | 0.3256 |
| Estimated Gini | 0.3078 | 0.3121 | 0.3309 | 0.3289 | 0.3235 | 0.3245 | 0.3308 | 0.3307 | 0.3416 | 0.3252 |
| Sample size | 8965 | 4873 | 8264 | 4502 | 8696 | 4674 | 11385 | 5865 | 11220 | 6046 |

TABLE 11

*Selected Dominance Probabilities for Metropolitan and Non-metropolitan Subgroups over Time*

|  | Metropolitan | Non-metropolitan |
|---|---|---|
| **2014 versus 2001** |  |  |
| $\Pr\left(2014 \succeq_{FSD} 2001\right)$ | 0.991 | 0.290 |
| $\Pr\left(2001 \succeq_{FSD} 2014\right)$ | 0.000 | 0.000 |
| $\Pr\left(\text{No FSD}\right)$ | 0.009 | 0.710 |
| $\Pr\left(2014 \succeq_{GLD} 2001\right)$ | 0.995 | 0.451 |
| $\Pr\left(2001 \succeq_{GLD} 2014\right)$ | 0.000 | 0.000 |
| $\Pr\left(\text{No GLD}\right)$ | 0.005 | 0.549 |
| **2010 versus 2001** |  |  |
| $\Pr\left(2010 \succeq_{FSD} 2001\right)$ | 0.878 | 0.580 |
| $\Pr\left(2001 \succeq_{FSD} 2010\right)$ | 0.000 | 0.000 |
| $\Pr\left(\text{No FSD}\right)$ | 0.122 | 0.420 |
| $\Pr\left(2010 \succeq_{GLD} 2001\right)$ | 0.951 | 0.708 |
| $\Pr\left(2001 \succeq_{GLD} 2010\right)$ | 0.000 | 0.000 |
| $\Pr\left(\text{No GLD}\right)$ | 0.049 | 0.292 |

TABLE 12

*Dominance probabilities for Metropolitan vs Non-metropolitan Subgroups*

|  | 2001 | 2006 | 2010 | 2014 | 2017 |
|---|---|---|---|---|---|
| $\Pr\left(\text{MET} \succeq_{FSD} \text{NON-MET}\right)$ | 0.2625 | 0.1562 | 0.6062 | 0.5386 | 0.1545 |
| $\Pr(\text{No FSD})$ | 0.7375 | 0.8438 | 0.3938 | 0.4614 | 0.8455 |
|  |  |  |  |  |  |
| $\Pr\left(\text{MET} \succeq_{GLD} \text{NON-MET}\right)$ | 0.5040 | 0.3727 | 0.8291 | 0.9923 | 0.2882 |
| $\Pr(\text{No GLD})$ | 0.4960 | 0.6273 | 0.1709 | 0.0077 | 0.7118 |
|  |  |  |  |  |  |
| $\Pr\left(\text{MET} \succeq_{LD} \text{NON-MET}\right)$ | 0.0058 | 0.0064 | 0.0078 | 0.0648 | 0.0006 |
| $\Pr\left(\text{NON-MET} \succeq_{LD} \text{MET}\right)$ | 0.0240 | 0.0280 | 0.0238 | 0.0000 | 0.0843 |
| $\Pr(\text{No LD})$ | 0.9702 | 0.9656 | 0.9684 | 0.9352 | 0.9151 |

*Note:* The omitted posterior probabilities for FSD and GLD of non-metropolitan over metropolitan for each year are all zero.

Table 13

*Restricted Dominance Probabilities for Metropolitan vs Non-metropolitan Subgroups for the Poorest 10% of the Population*

| | 2001 | 2006 | 2010 | 2014 | 2017 |
|---|---|---|---|---|---|
| $\Pr\left(\text{MET} \geq_{FSD} \text{NON-MET}\right)$ | 0.4120 | 0.3027 | 0.7169 | 0.9689 | 0.2397 |
| $\Pr(\text{No FSD})$ | 0.5880 | 0.6973 | 0.2831 | 0.0311 | 0.7603 |
| | | | | | |
| $\Pr\left(\text{MET} \geq_{GLD} \text{NON-MET}\right)$ | 0.5040 | 0.3727 | 0.8291 | 0.9923 | 0.2882 |
| $\Pr\left(\text{NON-MET} \geq_{GLD} \text{MET}\right)$ | 0.0033 | 0.0018 | 0.0000 | 0.0000 | 0.0001 |
| $\Pr(\text{No GLD})$ | 0.4927 | 0.6255 | 0.1709 | 0.0077 | 0.7117 |
| | | | | | |
| $\Pr\left(\text{MET} \geq_{LD} \text{NON-MET}\right)$ | 0.0109 | 0.0123 | 0.0290 | 0.2213 | 0.0093 |
| $\Pr\left(\text{NON-MET} \geq_{LD} \text{MET}\right)$ | 0.3578 | 0.5739 | 0.1115 | 0.0004 | 0.5323 |
| $\Pr(\text{No LD})$ | 0.6313 | 0.4138 | 0.8495 | 0.7783 | 0.4584 |

***Note:*** The omitted posterior probabilities for FSD of non-metropolitan over metropolitan for each year are all zero.

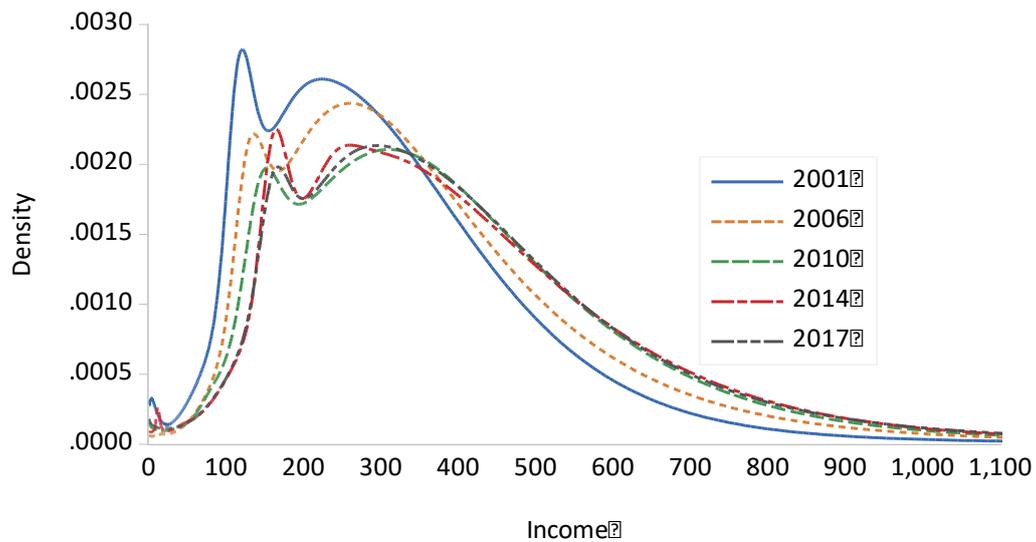

Figure 1

*Estimated Income Density Functions*

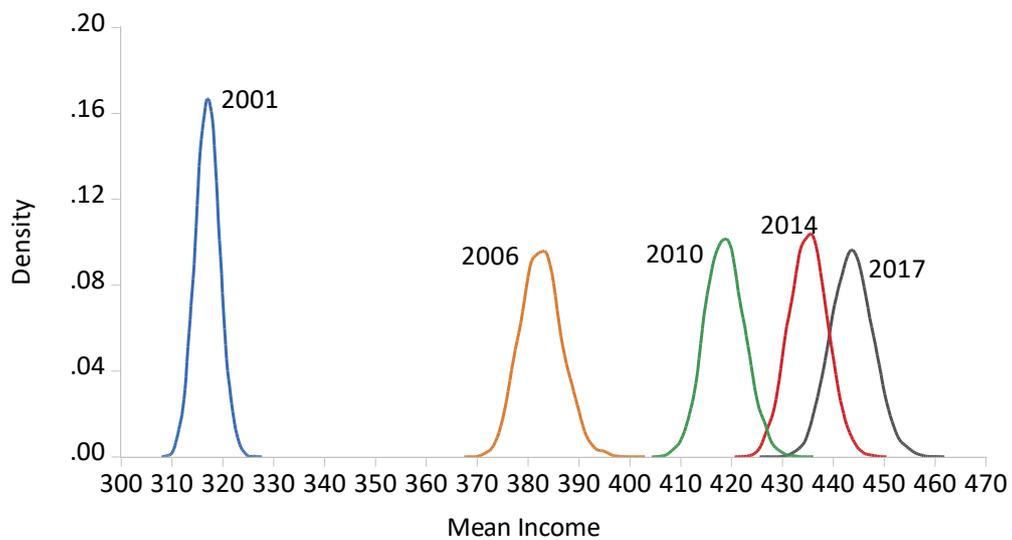

Figure 2

*Posterior Density Functions for Mean Incomes*

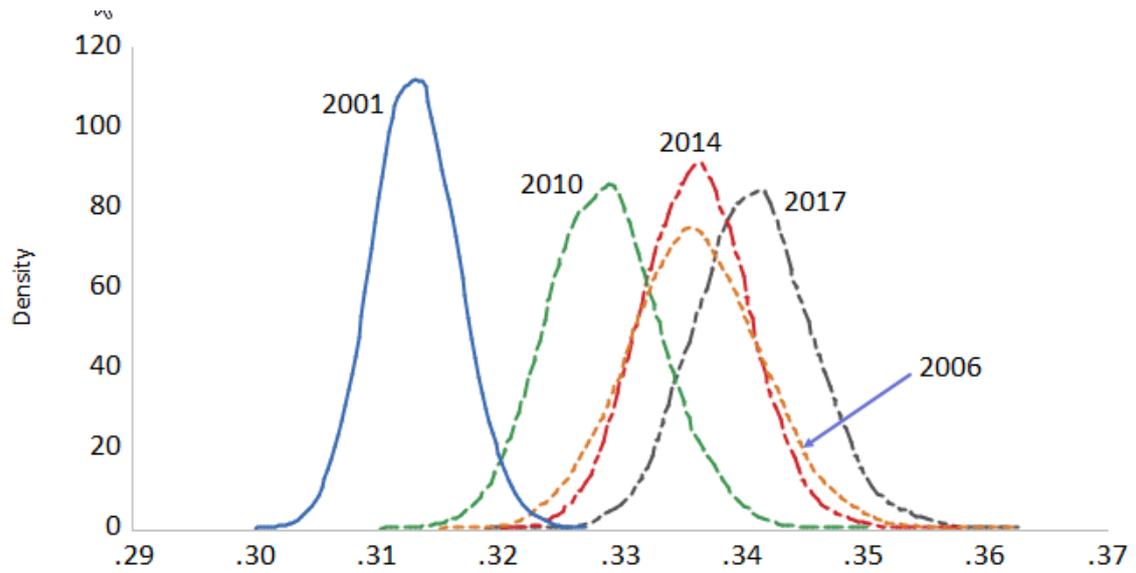

FIGURE 3

*Posterior Density Functions for Gini Coefficients*

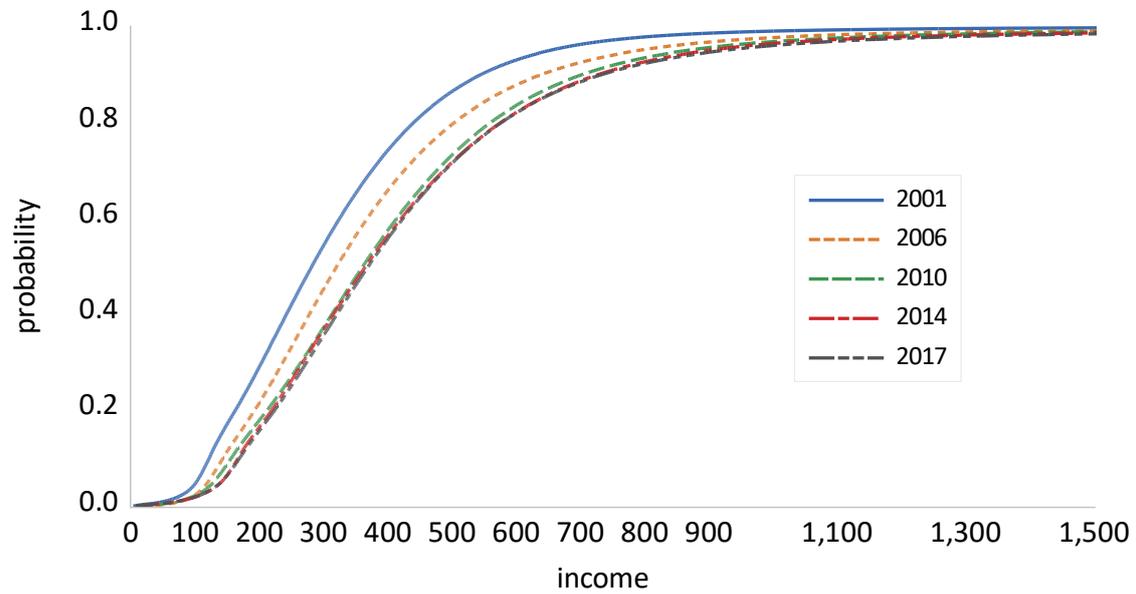

FIGURE 4

*Estimated Distribution Functions*

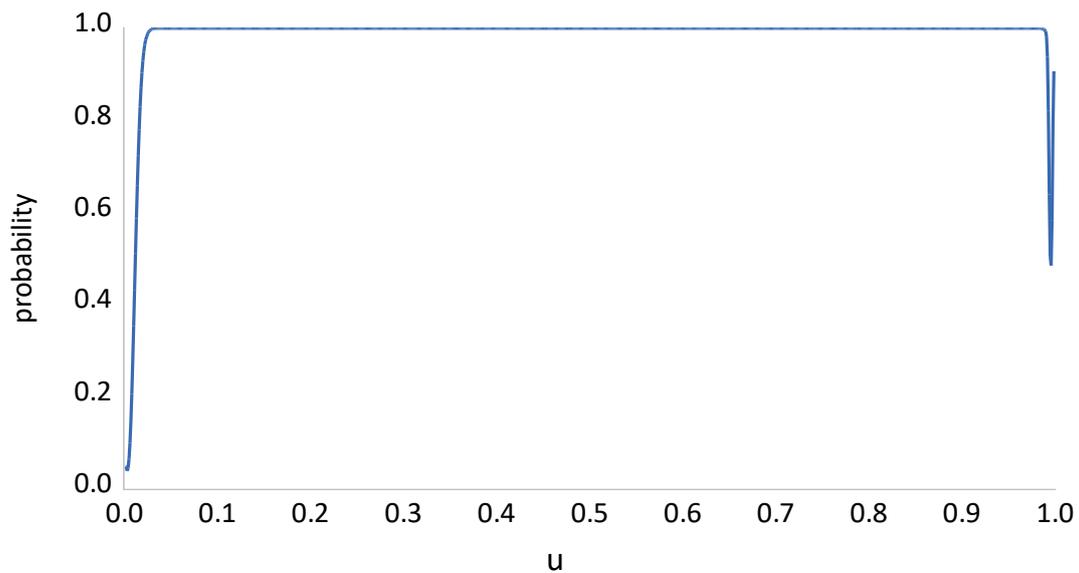

FIGURE 5

*Probability Curve for 2017 FSD 2006*

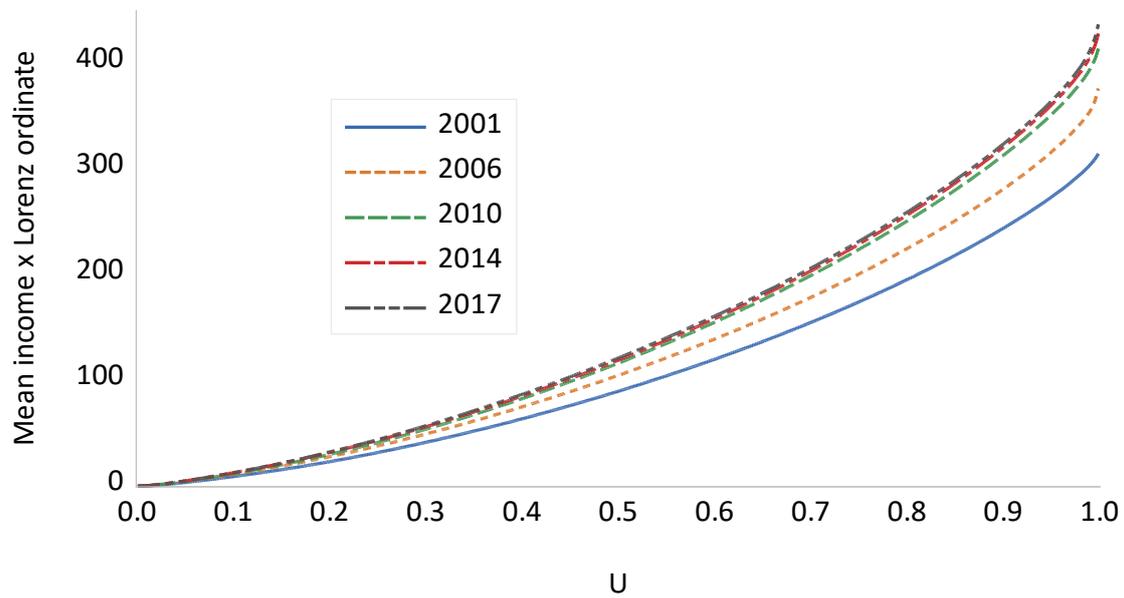

FIGURE 6

*Estimated Generalized Lorenz Curves*

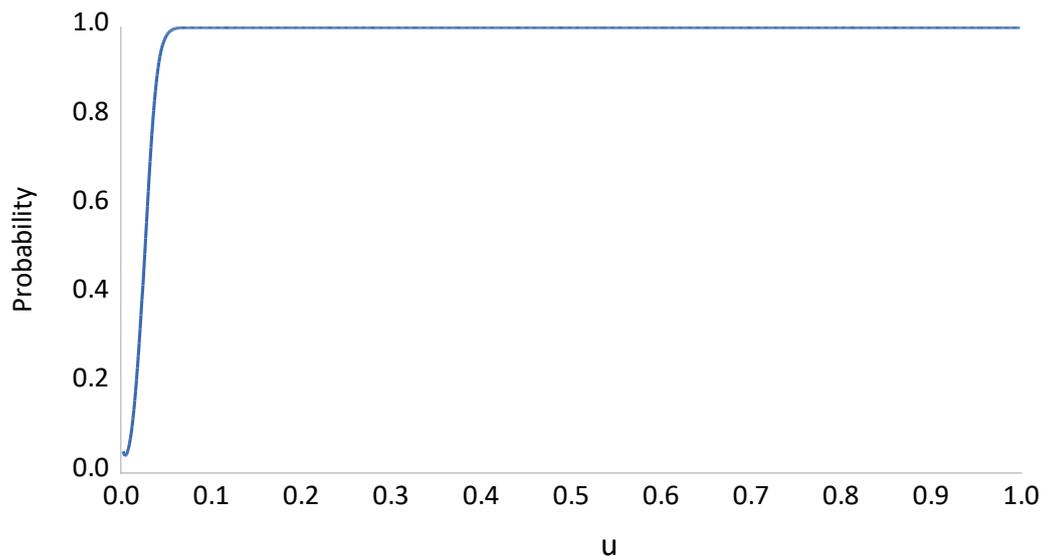

FIGURE 7

*Probability Curve for 2017 GLD 2006*

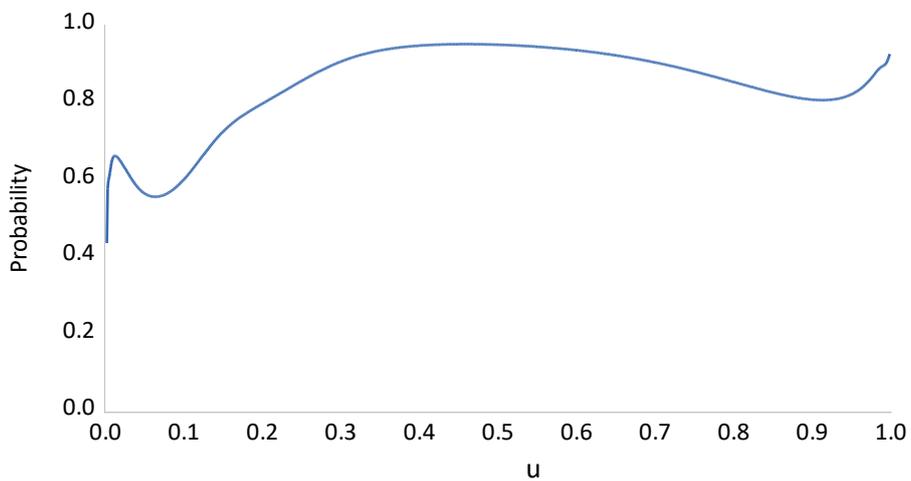

FIGURE 8

*Probability Curve for 2017 GLD 2014*

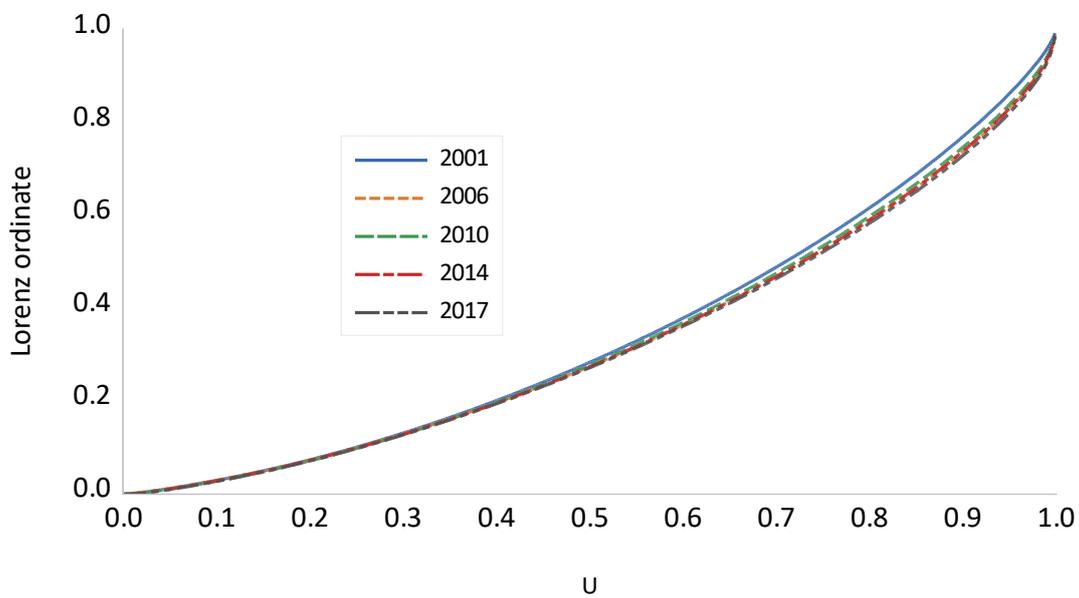

FIGURE 9

*Estimated Lorenz Curves*

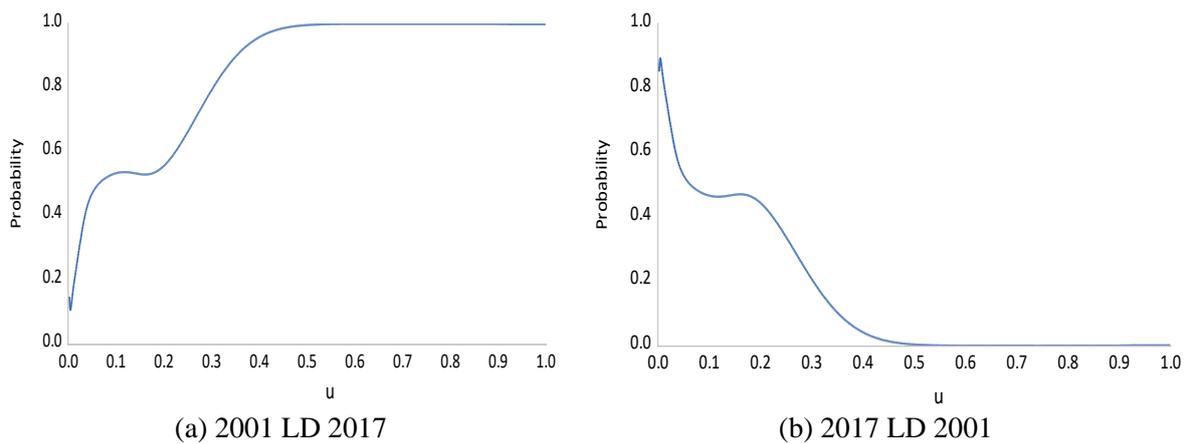

(a) 2001 LD 2017          (b) 2017 LD 2001

FIGURE 10

*Probability Curves for Lorenz Dominance: 2001 and 2017*

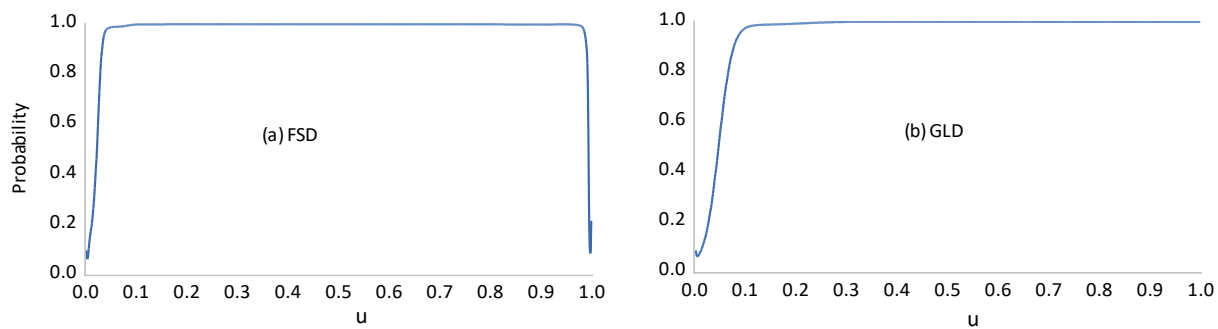

FIGURE 11

*Probability Curves for 2010 Dominating 2006*

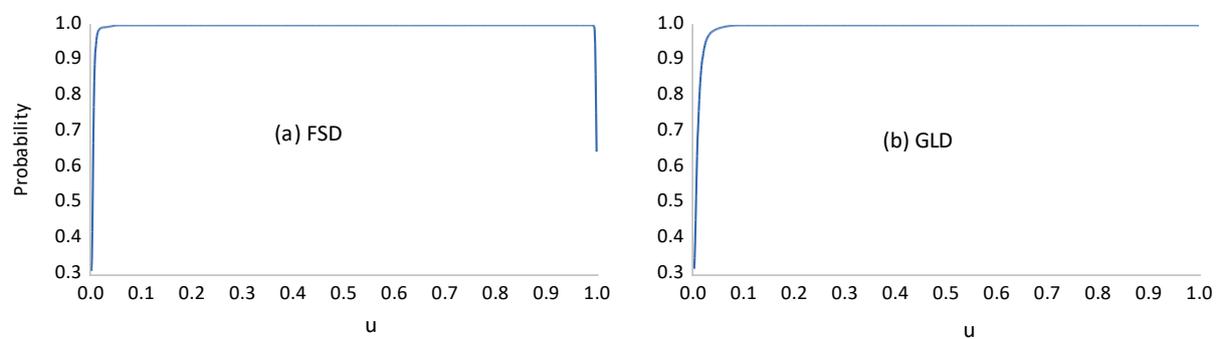

FIGURE 12

*Probability Curves for Metropolitan Dominating Non-metropolitan in 2017*

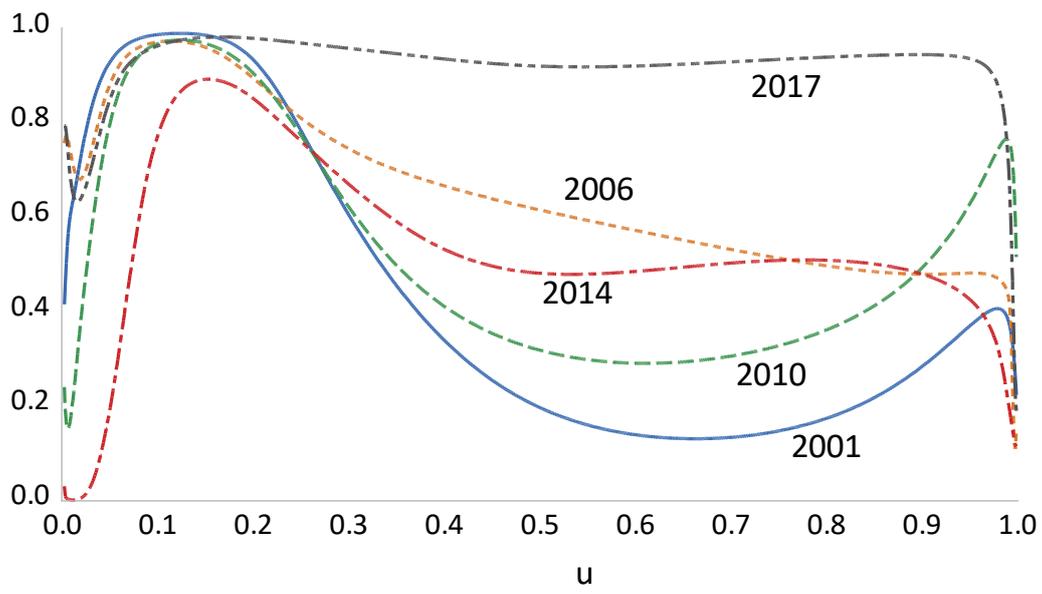

FIGURE 13

*Probability Curves for Non-metropolitan LD Metropolitan*